\documentclass[reprint,amsmath,amssymb,aps,showpacs,floatfix,byrevtex,longbibliography]{revtex4-1}

\pdfoutput=1
\usepackage{amsmath}
\usepackage{amssymb}
\usepackage{amstext}
\usepackage{amsopn}
\usepackage{amsfonts}
\usepackage{amsxtra}
\usepackage{color}
\usepackage{dcolumn}
\usepackage{graphicx}
\usepackage{hyperref}
\usepackage{bm}
\begin{document}

\preprint{Draft --- not for distribution}

%
%
\title{FeTe$_{\mathbf{0.55}}$Se$_{\mathbf{0.45}}$: a multiband superconductor in the clean and
  dirty limit}
\author{C. C. Homes}
\email{homes@bnl.gov}
\affiliation{Condensed Matter Physics and Materials Science Department,
  Brookhaven National Laboratory, Upton, New York 11973, USA}%
\author{Y. M. Dai}
\altaffiliation{Los Alamos National Laboratory, Center for Integrated
  Nanotechnologies, MPA-CINT, MS K771, Los Alamos, New Mexico 87545, USA}
\author{J. S. Wen}
\author{Z. J. Xu}
\author{G. D. Gu}
\affiliation{Condensed Matter Physics and Materials Science Department,
  Brookhaven National Laboratory, Upton, New York 11973, USA}%
\date{\today}

%
%
\begin{abstract}
The detailed optical properties of the multiband iron-chalcogenide superconductor FeTe$_{0.55}$Se$_{0.45}$
have been reexamined for a large number of temperatures above and below the critical temperature
$T_c=14$~K for light polarized in the \emph{a-b} planes.  Instead of the simple Drude model that assumes
a single band, above $T_c$ the normal-state optical properties are best described by the
two-Drude model that considers two separate electronic subsystems; we observe a weak response
($\omega_{p,D;1}\simeq 3000$~cm$^{-1}$) where the scattering rate has a strong temperature dependence
($1/\tau_{D,1}\simeq 32$~cm$^{-1}$ for $T \gtrsim T_c$), and a strong response ($\omega_{p,D;2}\simeq
14\,500$~cm$^{-1}$) with a large scattering rate ($1/\tau_{D,2}\simeq 1720$~cm$^{-1}$) that is
essentially temperature independent.  The multiband nature of this material precludes the use of
the popular generalized-Drude approach commonly applied to single-band materials, implying that any
structure observed in the frequency dependent scattering rate $1/\tau(\omega)$ is spurious and it
cannot be used as the foundation for optical inversion techniques to determine an electron-boson
spectral function $\alpha^2 F(\omega)$.
Below $T_c$ the optical conductivity is best described using two superconducting optical gaps of
$2\Delta_1\simeq 45$ and $2\Delta_2 \simeq 90$~cm$^{-1}$ applied to the strong and weak responses,
respectively.  The scattering rates for these two bands are vastly different at low temperature,
placing this material simultaneously in both clean and dirty limit.   Interestingly, this material
falls on the universal scaling line initially observed for the cuprate superconductors.

\end{abstract}
%
%
%
%
%
%
%
%
\maketitle

%
%
%
\section{Introduction}
The discovery of superconductivity in the iron-based materials \cite{kamihara08,ishida09,hsu08}
with maximum superconducting transition temperatures of $T_c \sim 55$~K achieved through
rare-earth substitution \cite{ren08} has prompted a tremendous amount of research into
the structural and electronic properties of this class of materials \cite{johnston10},
not only to ascertain the nature of the superconductivity but also to find a path to
higher transition temperatures.  Recently, attention has focused on the iron-chalcogenide
materials; these materials are structurally simple, consisting only of layers
of Fe$_2$(Se/Te)$_2$ tetrahedra \cite{hsu08}.
Nearly stoichiometric Fe$_{1+\delta}$Te undergoes a first-order magnetic and structural
transition \cite{bao09,zaliznyak11,zaliznyak12} from a tetragonal, paramagnetic state
to a monoclinic, antiferromagnetic state at $T_N \simeq 68$~K, but remains metallic
down to the lowest measured temperature.   Superconductivity has been observed at ambient
pressure in FeSe with $T_c = 8$~K \cite{hsu08}, increasing to $T_c \simeq 37$~K under
pressure \cite{medvedev09}.  The substitution of Te with Se in FeTe$_{1-x}$Se$_{x}$ suppresses
the structural and magnetic transition and establishes superconductivity over a broad range
of compositions\cite{liu10,zhuang14} with the critical temperature reaching a maximum
value \cite{fang08,chen_gf09,taen09,wu09,sales09,pourret11,dong11} of $T_c \simeq 14$~K for
$x \simeq 0.45$; enhanced $T_c$'s have been reported in thin films \cite{si09,huang10}.

Electronic structure calculations reveal a multiband material with three hole-like
bands at the origin and two electron-like bands at the corners of the Brillouin
zone \cite{subedi08}, a Fermi surface topology common to many of the iron-based
superconductors.  Angle resolved photoemission spectroscopy (ARPES) typically
identify most of these bands \cite{chen_f10,tamai10,lee10,ieki14,zhang14}.
Multiple isotropic superconducting energy gaps $\Delta \simeq 2 - 4$~meV have
been observed \cite{nakayama10,miao12}, and there is also evidence for
an anisotropic superconducting gap on one of the hole surfaces \cite{okazaki12}.
Despite being a multiband material with more than one type of free carrier, these
materials are poor metals \cite{sales09,pourret11}.
The optical properties in the Fe-Te/Se (\emph{a-b}) planes of FeTe$_{0.55}$Se$_{0.45}$
reveal a material that appears to be almost incoherent at room temperature but that develops
a metallic character just above $T_c$.  Below $T_c$ the emergence of a superconducting
state is seen clearly in the in-plane optical properties \cite{homes10,pimenov13,perucchi14}.
Perpendicular to the planes (\emph{c} axis) the transport appears incoherent and
displays little temperature dependence; below $T_c$ no evidence of a gap or a condensate
is observed \cite{moon11}.

%
%
In this work the detailed optical properties of FeTe$_{0.55}$Se$_{0.45}$ in the
\emph{a-b} planes are examined at a large number of temperatures in the normal
state and analyzed using the two-Drude model \cite{wu10a,heumen10}, which considers
two electronic subsystems rather than a single electronic band; this approach has
been successfully applied to thin films of this material \cite{perucchi14}.  The
single-band approach was used in a previous study of this material and was the basis
for the application of the generalized Drude model \cite{homes10}; however, we
demonstrate that the multiband nature of this material precludes the use of the
generalized Drude model.
The two-Drude model reveals a relatively weak Drude component ($\omega_{p,D;1}\simeq
3000$~cm$^{-1}$) with a small, strongly temperature dependent scattering rate at low
temperature ($1/\tau_{D,1}\simeq 32$~cm$^{-1}$), and a much stronger Drude component
where the strength ($\omega_{p,D;2} \simeq 14\,500$~cm$^{-1}$) and the scattering rate
($1/\tau_{D,2}\simeq 1720$~cm$^{-1}$) display little or no temperature dependence.
In the superconducting state the optical conductivity is reproduced quite well by
introducing isotropic superconducting gaps of $2\Delta_1 \simeq 45$~cm$^{-1}$ on the
broad Drude response, and $2\Delta_2 \simeq 90$~cm$^{-1}$ on the narrow Drude component;
no fitting is performed.  Comparing gaps and the scattering rates, we note that
$1/\tau_{D,1} \lesssim 2\Delta_1 (2\Delta_2)$, placing this close to the clean limit,
while $1/\tau_{D,2} \gg 2\Delta_1 (2\Delta_2)$, which is in the dirty limit; as a result,
this multiband material is simultaneously in both the clean and dirty limit.
The decomposition of the superconducting response into two bands allows the different
contributions to the superfluid density to be examined.  While the experimentally-determined
value and the clean-limit contribution falls on the universal scaling line for the
high-temperature superconductors \cite{homes04} in the region of the underdoped cuprates,
the dirty-limit contribution falls very close to the scaling line predicted for a dirty-limit
BCS superconductor \cite{homes05}.  New results for this scaling relation indicate that
it will be valid in both the clean and dirty limit \cite{kogan13}, which explains how this
material can satisfy both conditions and still fall on the scaling line.

%
%
\section{Experiment}
A mm-sized single crystal of FeTe$_{0.55}$Se$_{0.45}$ was cleaved from a piece of
the sample used in the original optical study\cite{homes10} revealing a flat, lustrous
surface along the Fe-Te/Se (\emph{a-b}) planes; this crystal has a critical temperature
of $T_c = 14$~K with a transition width of $\simeq 1$~K.
The reflectance has been measured at a near-normal angle of incidence for a large number
of temperatures (16) above and below $T_c$ over a wide frequency range ($\sim 3$~meV to
3~eV) for light polarized in the {\em a-b} planes using an {\em in situ} overcoating
technique \cite{homes93}.  The complex optical properties have been determined from a
Kramers-Kronig analysis of the reflectance \cite{dressel-book}, the details of which have
been previously described \cite{homes10}.

%
%
\section{Results and Discussion}
\subsection{Normal state}
\begin{figure}[t]
%
%
\centerline{\includegraphics[width=1.00\columnwidth]{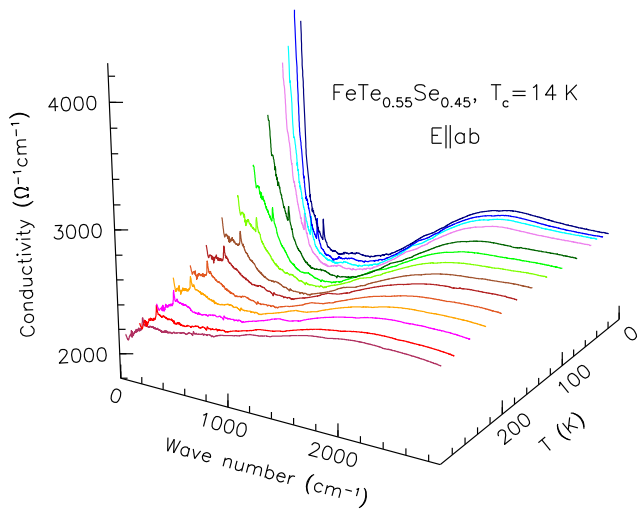}}%
\caption{(Color online) The real part of the in-plane optical conductivity of
FeTe$_{0.55}$Se$_{0.45}$ for a large number of temperatures in the normal state in the far
and mid-infrared region, showing the rapid emergence with decreasing temperature of
a Drude-like response at low frequency.}
\label{fig:water}
\end{figure}

%
%
The optical conductivity in the far and mid-infrared regions is shown for
a variety of temperatures above $T_c$ in the waterfall plot in Fig.~\ref{fig:water}.
At room temperature, the conductivity is essentially flat over the entire
frequency region.  The optical properties can be described using a simple
Drude-Lorentz model for the dielectric function:
\begin{equation}
  \tilde\epsilon(\omega) = \epsilon_\infty - {{\omega_{p,D}^2}\over{\omega^2+i\omega/\tau_{D}}}
    + \sum_j {{\Omega_j^2}\over{\omega_j^2 - \omega^2 - i\omega\gamma_j}},
\end{equation}
where $\epsilon_\infty$ is the real part at high frequency, $\omega_{p,D}^2 =
4\pi ne^2/m^\ast$ and $1/\tau_{D}$ are the square of the plasma frequency and
scattering rate for the delocalized (Drude) carriers, respectively, and $n$ and $m^\ast$
are the carrier concentration and effective mass.  In the summation, $\omega_j$,
$\gamma_j$ and $\Omega_j$ are the position, width, and strength of the $j$th vibration
or bound excitation.  The complex conductivity is $\tilde\sigma(\omega) = \sigma_1 +
i\sigma_2 = -i\omega [\tilde\epsilon(\omega) - \epsilon_\infty ]/60$ (in units of
$\Omega^{-1}$cm$^{-1}$).  The Drude response is simply a Lorentzian centered at
zero frequency with a full-width at half maximum of $1/\tau_D$.  The scattering
rate typically decreases with temperature, leading to a narrowing of the Drude
response and the transfer of spectral weight from high to low frequency, where
the spectral weight is the area under the conductivity curve, $N(\omega,T) =
\int_0^\omega \sigma_1(\omega^\prime)\,d\omega^\prime$.  As Fig.~\ref{fig:water}
indicates, while there is no clear free-carrier response at room temperature, there
is a rapid formation of a Drude-like response below about 200~K with a commensurate
transfer of spectral weight from high to low frequency below $\simeq 2000$~cm$^{-1}$.

%
%

%
%
%
The optical conductivity may be modeled quite well with only a single Drude term;
however, this is only possible if an extremely low-frequency Lorentzian oscillator
($\omega_0 \lesssim 3$~meV) is included.  While low-energy interband transitions are
expected for this class of materials \cite{valenzuela13}, they are not expected to
fall below $\simeq 30$~meV, well above the low-frequency oscillator required to fit
the data using this approach.
This suggests that a multiband system like FeTe$_{0.55}$Se$_{0.45}$ is more
correctly described by a two-Drude model \cite{wu10a,heumen10} in which the electronic
response is modeled as two separate, uncorrelated electronic subsystems rather
than a single dominant band.  Using this approach, the second term in Eq.~(1)
becomes a summation in which the plasma frequency and the scattering rate are
now indexed over the total number of bands under consideration (two in this
case).  Both the real and imaginary parts of the conductivity are
fit simultaneously using a non-linear least-squares method, which allows very
broad features to be fit more reliably than fitting to just the real part of
the optical conductivity alone.

%
%
\begin{figure}[t]
%
%
\centerline{\includegraphics[width=0.95\columnwidth]{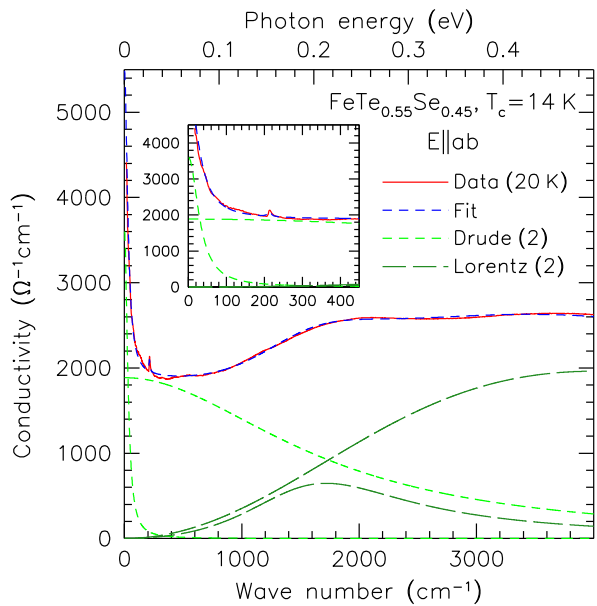}}%
\caption{(Color online) The Drude-Lorentz model fit to the real part
of the optical conductivity of FeTe$_{0.55}$Se$_{0.45}$ at 20~K for light
polarized in the \emph{a-b} planes for two Drude components and two Lorentz
oscillators.  The inset shows the linear combination of the two Drude
components in the low frequency region; the sharp structure at $\simeq
204$~cm$^{-1}$ is the normally infrared-active $E_u$ mode \cite{homes10}. }
\label{fig:fits}
\end{figure}
The result of the fit to the data at 20~K is shown in Fig.~\ref{fig:fits}, revealing
two distinct Drude components; a narrow response with $\omega_{p,D;1}\simeq 2630$~cm$^{-1}$
and $1/\tau_{D,1}\simeq 32$~cm$^{-1}$, and a much broader and stronger component with
$\omega_{p,D;2}\simeq 14\,110$~cm$^{-1}$ and $1/\tau_{D,2} \simeq 1770$~cm$^{-1}$.
These values are consistent with the results from the two-Drude analysis performed
on FeTe$_{0.5}$Se$_{0.5}$ thin films \cite{perucchi14}.  The structure in the mid-infrared
region is described by two oscillators centered at $\omega_1 \simeq 1720$ cm$^{-1}$ and
$\omega_2 \simeq 4010$~cm$^{-1}$; other high-frequency oscillators have been included to
describe the optical conductivity in the near-infrared and visible regions, but they are
not shown in this plot.

%
%
\begin{figure}[b]
%
%
\centerline{\includegraphics[width=0.90\columnwidth]{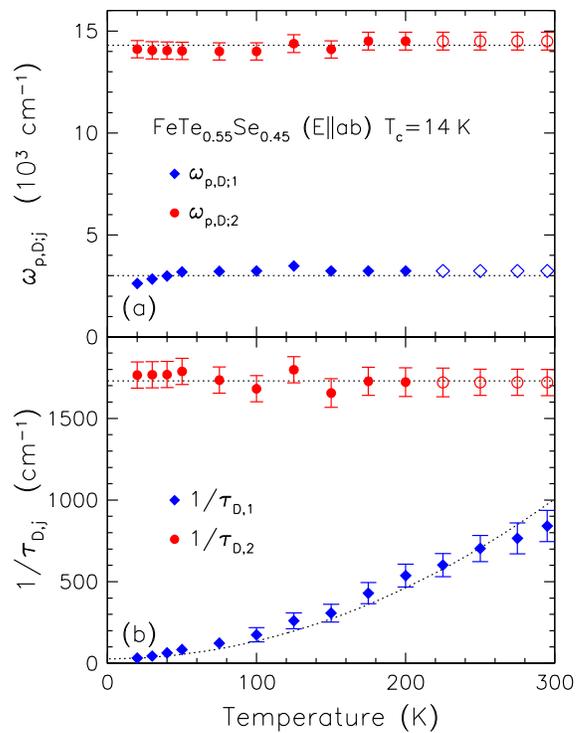}}%
\caption{(Color online) The two-Drude model fit to the optical conductivity
yielding the temperature dependence of the (a) plasma frequencies $\omega_{p,D;j}$
and (b) scattering rates $1/\tau_{D,j}$ for the narrow (diamonds) and broad
(circles) Drude components in FeTe$_{0.55}$Se$_{0.45}$ for $T>T_c$.  The
filled symbols indicate fitted parameters, while the open symbols indicate
that the parameter held fixed to a constant value.  Where error bars are not
shown, the error is roughly the size of the symbol.  The dotted lines are
drawn as a guide for the eye.}
\label{fig:drude}
\end{figure}

It is not immediately obvious if the narrow Drude component originates from
the electron or the hole pockets.  In a previous study of the non-superconducting
parent compound Fe$_{1.03}$Te, the weak Drude-like feature at high temperature
and the the rapid increase of the low-frequency conductivity below the magnetic
and structural transition at $T_N \simeq 68$~K was associated with the closing of
the pseudogap on the electron pocket \cite{lin13,dai14}.  While the scattering
rate on the electron pocket in Fe$_{1.03}$Te was observed to be about 6~meV at
low temperature, it also displayed relatively little temperature dependence,
whereas in the current study the pocket with the a scattering rate about 4~meV
at 20~K shows considerable temperature dependence.  In ARPES studies of iron-arsenic
superconductors, small scattering rates ($\simeq 3$~meV) have been observed on both
the electron and hole pockets at low temperature \cite{umezawa12}, which is consistent
with the observation that electron and hole mobilities are similar at low
temperature in FeTe$_{0.5}$Se$_{0.5}$, unlike Fe$_{1+\delta}$Te where the electron
mobility is much larger than that of the holes below $T_N$ (Ref.~\onlinecite{tsukada11}).
While it is tempting to associate the small scattering rate with an electron pocket,
we can not make any definitive statements at this point.

%
%
\begin{figure}[t]
%
%
\centerline{\includegraphics[width=0.95\columnwidth]{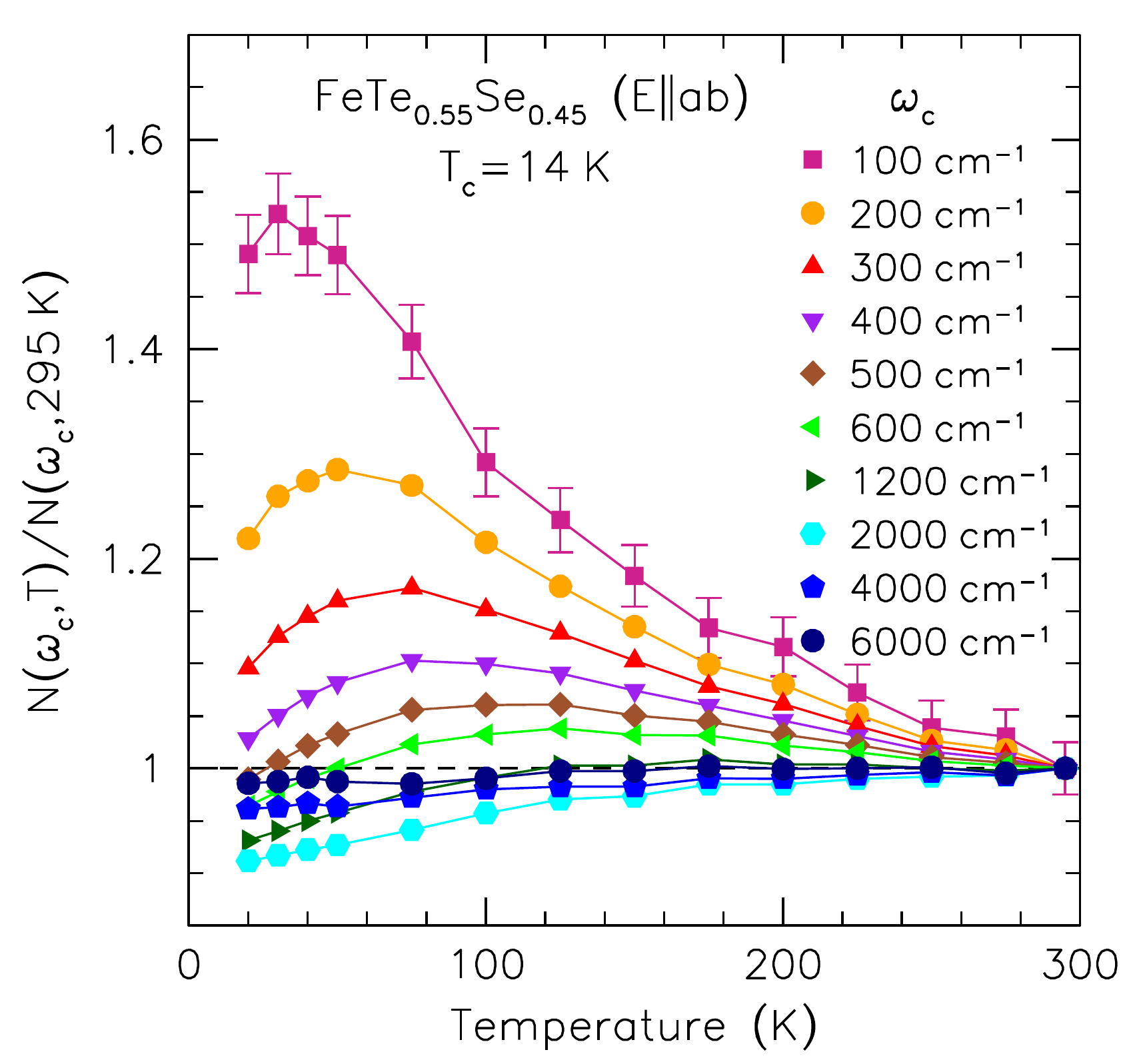}}%
\caption{(Color online) The temperature dependence of the spectral
weight normalized to the value at 295\,K for a variety of choices
for the cut-off frequency $\omega_c$; the estimated error is indicated
for the $\omega_c = 100$~cm$^{-1}$ points.  Smaller values of $\omega_c$
result in a strong temperature dependence; however, within the
confidence limits of the experiment, for $\omega_c \gtrsim 4000$~cm$^{-1}$
there is effectively little or no temperature dependence. }
\label{fig:weight}
\end{figure}

The two-Drude model has been used to fit the real and imaginary parts of
the optical conductivity in the normal state for $T > T_c$; the temperature
dependence of the plasma frequencies and the scattering rates for the narrow
and broad components are shown in Fig.~\ref{fig:drude}.  The fit to the optical
conductivity at 20~K, and at low temperatures in general, is unambiguous due
to the narrow Drude term; as a result, both the plasma frequencies and the
scattering rates may be fit simultaneously.  As Fig.~\ref{fig:drude}(a)
indicates, at low temperature the plasma frequency for the broad component
displays little temperature dependence, while the plasma frequency for the
narrow Drude component decreases slightly just above $T_c$.
At low temperatures, the scattering rate for the broad component also displays
little temperature dependence, whereas the scattering rate for the narrow
component increases quickly with temperature, until by 200~K it has
increased by a factor of $\simeq 20$.  At high temperature, the presence of
two broad Drude terms makes the fit to the now relatively featureless complex
conductivity more challenging.  As a result, above 200~K the fit is constrained
to only the scattering rate for the narrow Drude term; both plasma frequencies
and the scattering rate for the broad Drude term are held fixed.  This is
indicated in Fig.~\ref{fig:drude} by the solid symbols (fitted parameters),
and the open symbols (fixed parameters).
Using these constraints, the scattering rate for the narrow Drude term continues
to increase until at room temperature $1/\tau_{D,1} \simeq 840$~cm$^{-1}$, about
half the value of the scattering rate observed for the other Drude component.
The dotted line shown in Fig.~\ref{fig:drude} for $1/\tau_{D,1}$ has the quadratic
form that would be expected for a Fermi liquid; however, below 100~K the data
may be fit equally well by a straight line, making it difficult to draw
any conclusions about the nature of the transport on this pocket.

%
%
Returning to the evolution of the conductivity in the normal state, it is
clear from Fig.~\ref{fig:water} that the growth of the low-frequency Drude
component is accompanied by the loss of spectral weight throughout much of
the infrared region; however, it is important note that these changes occur
on top of a large background conductivity that originates from the strong
Drude component and several mid-infrared absorptions.  To estimate the energy
scale over which this transfer takes place, the normalized spectral weight
$N(\omega_c,T)/N(\omega_c,295\,{\rm K})$  is plotted in Fig.~\ref{fig:weight}
for a variety of choices for the cut-off frequency, $\omega_c$.  Small values
of $\omega_c$ result in a strong temperature dependence.  Normally, larger
values of $\omega_c$ would eventually result in a temperature-independent
curve with a value of unity; however, before this occurs the ratio is first
observed to drop below unity for $\omega_c \simeq 600$~cm$^{-1}$ before
finally adopting the expected form for $\omega_c \gtrsim 4000$~cm$^{-1}$.
We speculate that this is in response to the reduction of the plasma frequency
of the narrow Drude component at low temperature resulting in a transfer of
spectral weight from a coherent to an incoherent response at high frequency.
This effect has in fact been predicted in the iron-based materials and is
attributed to Fermi surface reduction due to many body effects \cite{benfatto11b}.
Finally, we remark that while the redistribution of spectral weight in the
parent compound Fe$_{1.03}$Te below $T_N$ is due to the closing of the pseudogap
on the electron pocket in that material \cite{lin13,dai14}, in the present case it
is due to the slight decrease in the plasma frequency and the dramatic decrease in
the scattering rate of the narrow Drude component at low temperature.

%
%
%
\subsection{Generalized Drude model}
Beyond the two-component Drude-Lorentz and the two-Drude approaches for
modeling the optical conductivity, there is a third approach, the generalized
Drude model.  This latter approach is commonly used to describe the normal
state of the cuprate materials where only a single band crosses the Fermi
level, and is referred to as a single component model.  The optical conductivity
of the cuprates is similar to that of FeTe$_{0.55}$Se$_{0.45}$; typically,
just above $T_c$, there is a narrow Drude-like response that gives way to
a flat, incoherent mid-infrared component, resulting in a kink-like feature
in the optical conductivity \cite{thomas88,tu02,basov05}.  This kink is attributed
to a strongly-renormalized scattering rate due to electron-boson coupling, and
is is described in the generalized Drude model through a frequency-dependent
scattering rate and effective mass \cite{allen77,puchkov96},
\begin{equation}
  {{1}\over{\tau(\omega)}} = {{\omega_p^2}\over{4\pi}} \,
  {\rm Re} \left[ {{1}\over{\tilde\sigma(\omega)}} \right]
  \label{eq:tau}
\end{equation}
and
\begin{equation}
  {{m^\ast(\omega)}\over{m_e}} = {{\omega_p^2}\over{4\pi\omega}} \,
  {\rm Im} \left[ {{1}\over{\tilde\sigma(\omega)}} \right] ,
\end{equation}
where $m_e$ is the bare mass, $m^\ast(\omega)/m_e = 1+\lambda(\omega)$
and $\lambda(\omega)$ is a frequency-dependent electron-boson
coupling constant.  The frequency-dependent scattering rate is
the basis for optical inversion methods to calculate the electron-boson spectral
function \cite{carbotte99,dordevic05}.
%
%
\begin{figure}[t]
%
%
\centerline{\includegraphics[width=0.90\columnwidth]{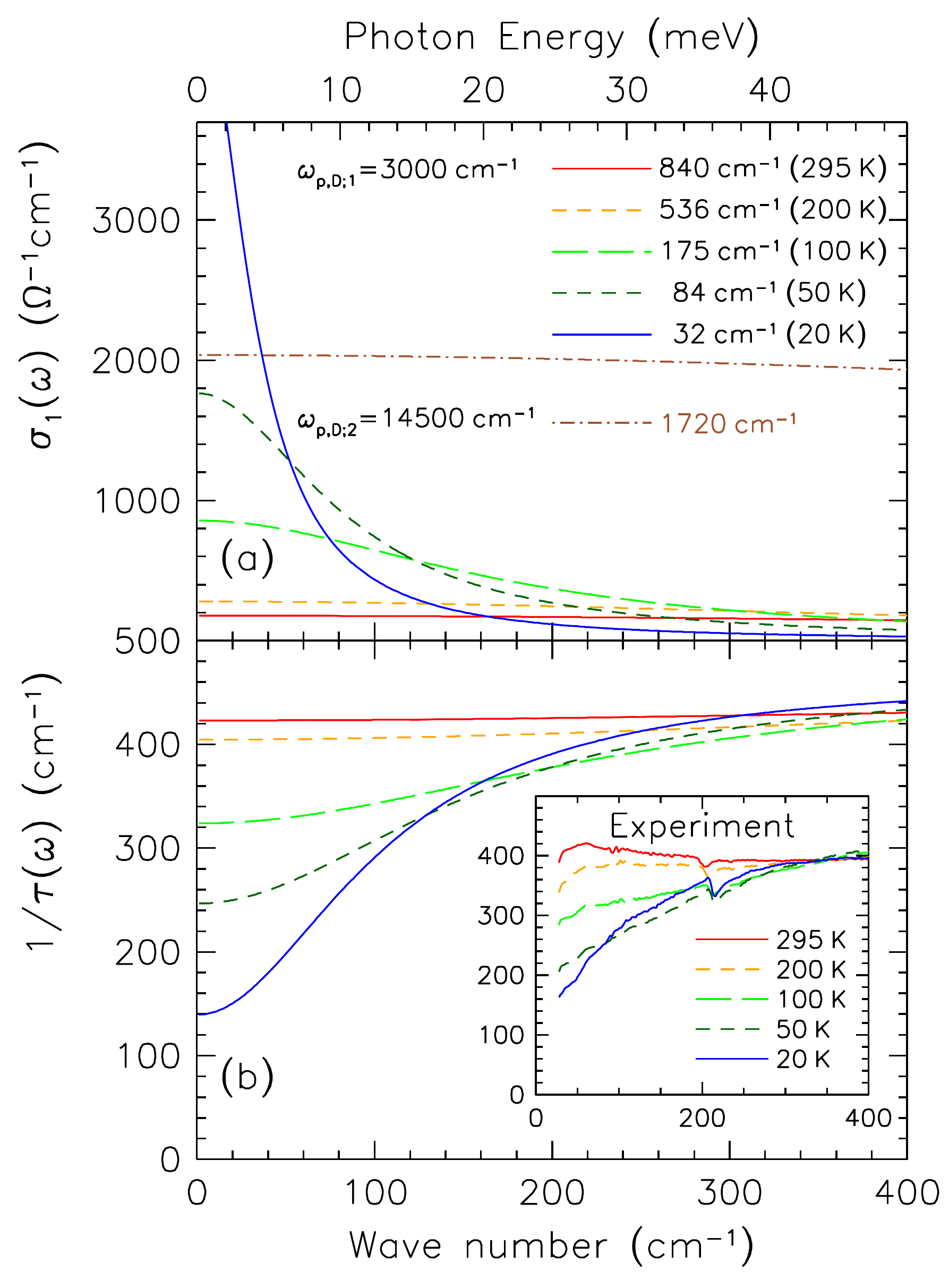}}%
\caption{(Color online) (a) The optical conductivity for the temperature-independent
broad, strong Drude component (dot-dash line), and the weaker Drude component (solid and
dashed lines) that displays a strongly temperature-dependent scattering rate.
(b) The frequency-dependent scattering rate calculated from the two-Drude
model.  Inset: the experimentally-determined $1/\tau(\omega)$.}
\label{fig:general}
\end{figure}
However, concerns have been raised over the effect of the low-energy
interband transitions on the scattering rate \cite{benfatto11a}, and more
generally, the multiband nature of the iron pnictide and iron selenide
materials presents a major difficulty for this type of analysis. To
illustrate this point, we consider the complex dielectric function for
a two Drude model.  The plasma frequency for the weak component has
been taken to be with $\omega_{p,D;1}\simeq 3000$~cm$^{-1}$; initially
the scattering rate is quite broad with $1/\tau_{D,1}\simeq 840$~cm$^{-1}$
at 295~K, but as Fig.~\ref{fig:drude} indicates it decreases rapidly
with temperature to $1/\tau_{D;1}\simeq 32$~cm$^{-1}$ at 20~K.  The optical
conductivity at these temperatures, as well as 200, 100, and 50~K, are shown
in Fig.~\ref{fig:general}(a) as the various lines.  In addition, a broad, strong
Drude component with $\omega_{p,D;2}\simeq 14\,500$~cm$^{-1}$ and $1/\tau_{D,2}
\simeq 1720$~cm$^{-1}$ is shown in Fig.~\ref{fig:general}(a) as a dash-dot
line; this component is temperature independent.  From these two Drude
responses a temperature-dependent complex dielectric function is constructed and
the frequency-dependent scattering rate is calculated from Eq.~(\ref{eq:tau})
using a somewhat arbitrary value of $\omega_p \simeq 7500$~cm$^{-1}$; the
result is shown in Fig.~\ref{fig:general}(b).  The actual experimental values
are shown in the inset using values of $\omega_p \simeq 6700 - 7300$~cm$^{-1}$,
where $\omega_p$ has been chosen so that the values for the scattering rate are
roughly the same at 400~cm$^{-1}$.
At 295~K, where the scattering rates are broad, $1/\tau(\omega)$ displays
little or no frequency dependence, and the same can be said of the result
at 200~K; this type of response would be expected from a simple Drude model
with only a single component.  This trend does not continue; by 100~K
the scattering rate has developed strong frequency dependence and by
20~K the scattering rate has a linear frequency dependence over much of the
low-frequency region.  In a previous single-component analysis
of this material, this $1/\tau(\omega) \propto \omega$ behavior was taken as
evidence for electronic-correlations \cite{homes10}.  However, the multiband
nature of this material indicates that the linear-frequency dependence
observed in $1/\tau(\omega)$ is simply a consequence of having more than
one Drude component.  As a result, unless the system has been heavily doped
into a regime where it is either purely electron or hole doped, then the
single-component, generalized-Drude approach should be avoided.  It should
also not be used as a basis for optical-inversion techniques used to calculate
the electron-boson spectral function.

%
%
\subsection{Superconducting state}
\subsubsection{Superfluid density}
\begin{figure}[b]
%
%
\centerline{\includegraphics[width=0.90\columnwidth]{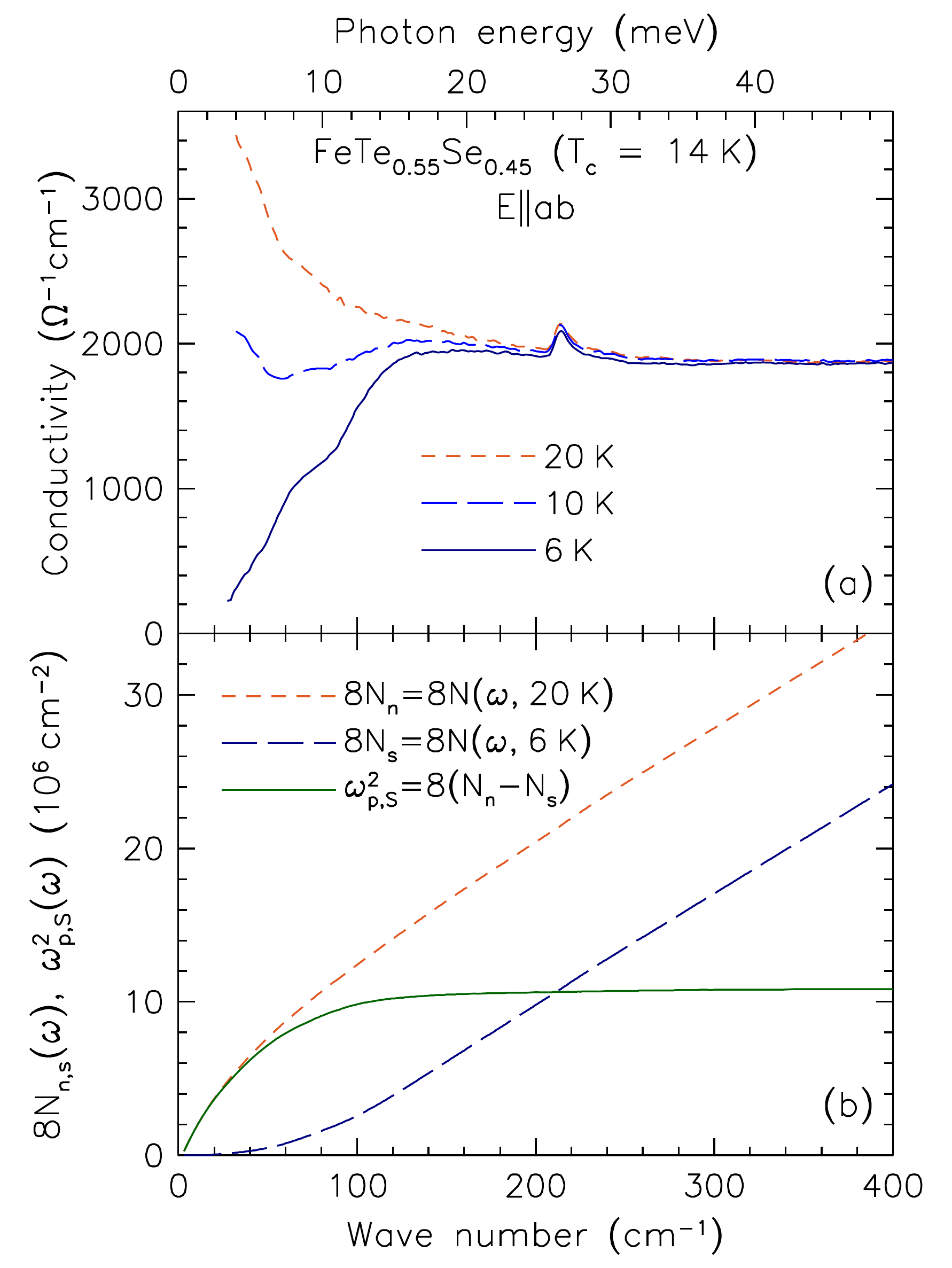}}%
\caption{(Color online) (a) The real part of the optical conductivity for
FeTe$_{0.55}$Se$_{0.45}$ for light polarized in the \emph{a-b} planes just
above $T_c$ at 20~K, and at two temperatures below $T_c$.  Note the strong
suppression of the low-frequency conductivity for $T\ll T_c$.
(b) The spectral weight in the normal state, $N_n(\omega, T\gtrsim T_c)$
and in the superconducting state, $N_s(\omega, T\ll T_c)$; the expression
for $\omega_{p,S}^2$ converges by $\omega \simeq 150$~cm$^{-1}$.}
\label{fig:fgt}
\end{figure}

While the optical conductivity in the normal state in Fig.~\ref{fig:water}
shows the development of a strong Drude-like component at low temperature, upon
entry into the superconducting state there is a dramatic suppression of the
low-frequency conductivity and a commensurate loss of spectral weight, shown
in Fig.~\ref{fig:fgt}(a).  The loss of spectral weight is associated with
the formation of a superconducting condensate, whose strength may be calculated
from the Ferrell-Glover-Tinkham (FGT) sum rule \cite{ferrell58,tinkham59}:
\begin{equation}
  \int_{0^+}^{\omega_c} \left[ \sigma_1(\omega, T \gtrsim T_c) -
  \sigma_1(\omega, T\ll T_c) \right] d\omega = \omega_{p,S}^2/8,
  \label{eq:fgt}
\end{equation}
or $\omega_{p,S}^2=8\left[ N_n(\omega_c,T\gtrsim T_c) -
N_s(\omega_c,T\ll T_c)\right]$, where $\omega_c$ is chosen so that the integral
converges and $\omega_{p,S}^2 = 4\pi n_s e^2/m^*$ is the superconducting
plasma frequency.  The superfluid density is $\rho_{s0} \equiv \omega_{p,S}^2$.
The evolution of the spectral weight for $N_n$ and $N_s$ are shown in Fig~\ref{fig:fgt}(b).
It is apparent from Fig.~\ref{fig:fgt}(a) that most of the changes in the spectral
weight occur below $\simeq 100$~cm$^{-1}$, so it is therefore not surprising that
the expression for $\omega_{p,S}^2$ has converged for $\omega_c \simeq 120$~cm$^{-1}$.
%
%
The sum rule yields $\omega_{p,S} \simeq 3280\pm 200$~cm$^{-1}$, from which
an effective penetration depth can be calculated, $\lambda_0 = 4850\pm 300$~\AA ,
slightly smaller than the result obtained in the previous optical study \cite{homes10},
and in good agreement with values of $\lambda_0 \simeq 4300 - 5600$~\AA\ observed in
materials with similar composition measured using several different
methods \cite{kim10,biswas10,klein10,takahashi11}.
In a previous single-band interpretation of the optical conductivity of this material,
it was noted that $\omega_{p,S} \ll \omega_{p,D}$, suggesting that only small
portion of the free carriers collapsed into the condensate below $T_c$ and that
this material was therefore not in the clean limit.  However, the multiband
nature of this compound results in a more complicated picture where this
statement is only partially true.

%
%
\subsubsection{Multiband superconductor}
\begin{figure}[t]
%
%
\centerline{\includegraphics[width=0.90\columnwidth]{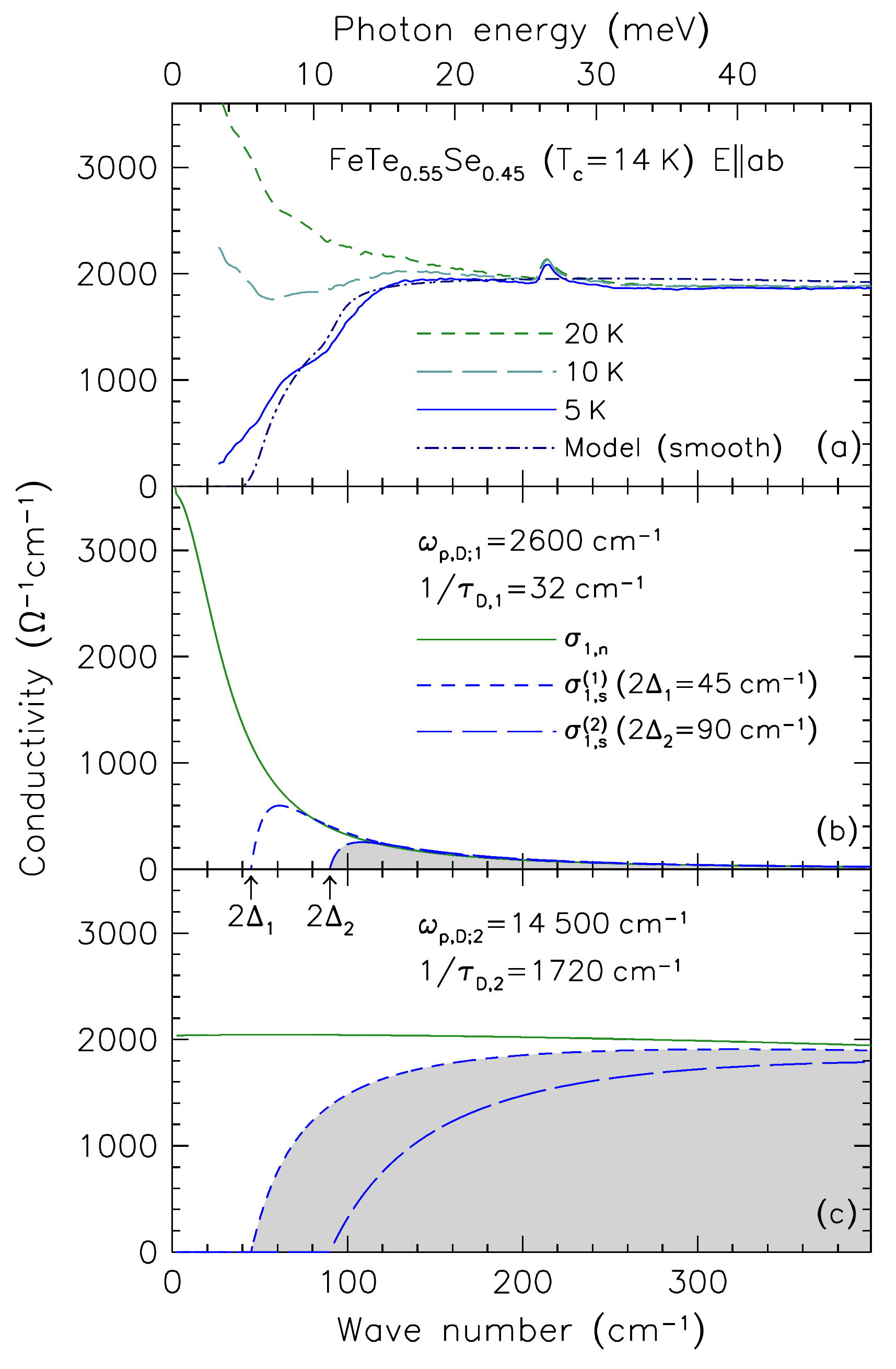}}%
\caption{(Color online) (a) The real part of the optical conductivity for
FeTe$_{0.55}$Se$_{0.45}$ for light polarized in the \emph{a-b} planes just
above $T_c$ at 20~K, and at two temperatures below $T_c$.  The dash-dot
line models the smoothed contribution to the conductivity from the gapped
excitations described in the discussion.
(b) The real part of the optical conductivity for a Drude model with
$\omega_{p,D;1} = 2600$~cm$^{-1}$ and $1/\tau_{D,1}=32$~cm$^{-1}$ (solid line),
and the contribution from the gapped excitations for $T\ll T_c$ with
superconducting gaps of $2\Delta_1=45$~cm$^{-1}$ and $2\Delta_2=90$~cm$^{-1}$
(dashed lines).
(c) The same set of calculations for $\omega_{p,D;2} = 14\,500$~cm$^{-1}$
and $1/\tau_{D,2}=1720$~cm$^{-1}$.}
\label{fig:bcs}
\end{figure}

The complex optical conductivity shown in Fig.~\ref{fig:fgt}(a) is reproduced
in Fig.~\ref{fig:bcs}(a); as previously noted, below $T_c$ most of the transfer
of spectral weight occurs below $\simeq 120$~cm$^{-1}$, setting a na\"{\i}ve energy
scale for the maximum value of the superconducting energy gap.  In addition
to the general suppression of the optical conductivity below 120~cm$^{-1}$, there
is also a shoulder at $\simeq 60$~cm$^{-1}$, suggesting more than one energy
scale for superconductivity in this material \cite{tu01}.  In the previous work
where a single-band interpretation was employed \cite{homes10}, the optical
conductivity was reproduced reasonably well by using a Mattis-Bardeen formalism
for the contribution from the gapped excitations \cite{zimmerman91,dressel-book}.
The Mattis-Bardeen approach assumes that $l\lesssim \xi_0$, where the
mean-free path $l = v_F \tau$ ($v_F$ is the Fermi velocity), and the coherence
length is $\xi_0 = \hbar v_F/\pi\Delta_0$ for an isotropic superconducting gap
$\Delta_0$; this may also be expressed as $1/\tau \gtrsim 2\Delta_0$.
The best result was obtained by using two isotropic superconducting energy
gaps of $2\Delta_1=40$~cm$^{-1}$ and $2\Delta_2=83$~cm$^{-1}$, where a
moderate amount of disorder-induced scattering was introduced \cite{homes10}.
However, in the two-Drude model, the amount of scattering in each band is
dramatically different, $1/\tau_{D,1}\ll 1/\tau_{D,2}$.  To model the data,
we use the values for the plasma frequencies and the scattering rates just
above $T_c$ at 20~K, shown in Fig.~\ref{fig:drude}, for the two different bands;
the two isotropic superconducting energy gaps are taken to be $2\Delta_1 =
45$~cm$^{-1}$ and $2\Delta_2 = 90$~cm$^{-1}$.  The contribution from each
of the gapped excitations is then calculated.   We emphasize at this point
that no fitting is employed and that the parameters are not refined.

%
%
The solid line in Fig.~\ref{fig:bcs}(b) shows the normal-state conductivity
for $\omega_{p,D;1}=2600$~cm$^{-1}$ and $1/\tau_{D,1}=32$~cm$^{-1}$ for $T\gtrsim
T_c$, while dashed lines denote the contributions from the gapped excitations
from $2\Delta_1$ and $2\Delta_2$ for $T\ll T_c$.  Below the superconducting energy
gap the conductivity is zero and there is no absorption, while above the gap
there is a rapid onset of the conductivity, which then joins the normal-state value
at higher energies.  Using the FGT sum rule in Eq.~(\ref{eq:fgt}) we estimate
$\omega_{p,S} \simeq 2150$~cm$^{-1}$ for the lower gap and $\omega_{p,S} \simeq
2300$~cm$^{-1}$ for the upper gap, indicating that about $70-80$\% of the free
carriers collapse into the condensate for $T\ll T_c$.  This is consistent with
the observation that $1/\tau_{D,1} \lesssim 2\Delta_1, 2\Delta_2$, placing this
material in the moderately-clean limit.  It has been remarked that for a
single-band material in the clean limit the opening of a superconducting energy
gap may be difficult to observe because the small normal-state scattering rate
can lead to a reflectance that is already close to unity, thus the increase in
the reflectance below $T_c$ for $\omega \lesssim 2\Delta$ is difficult to
observe \cite{kamaras90}.  However, this is a multiband material in which the
overall superconducting response arises from the gapping of several bands,
some of which are not necessarily in the clean limit, discussed below.

%
%
%

%
%
The same procedure is carried out for the second band in Fig.~\ref{fig:bcs}(c)
for $\omega_{p,D;2}=14\,500$~cm$^{-1}$ and $1/\tau_{D,2}=1720$~cm$^{-1}$.  Here,
the normal-state conductivity is nearly flat in the low-frequency region.  For
$T \ll T_c$, the conductivity is once again zero below the superconducting energy
gap; however, unlike the previous case the onset of conductivity above the gap now
takes place much more slowly.  In addition, the curves only merge with the normal-state
values at energies well above the values for the superconducting gaps.  From the
FGT sum rule, we estimate $\omega_{p,S} \simeq 2740$~cm$^{-1}$ for the lower gap
and $\omega_{p,S} \simeq 3670$~cm$^{-1}$ for the upper gap, indicating that about
$3-6$\% of the free carriers collapse into the condensate for $T\ll T_c$.  This
is consistent with the observation that $1/\tau_{D,2} \gg 2\Delta_1, 2\Delta_2$,
placing this material in the dirty limit.
%
%
Thus, as a consequence of the multiband nature of this material, it can
coexist in both the clean and dirty limit at the same time; we speculate
that this condition is likely fulfilled in many (if not all) of the iron-based
superconductors.

While we have considered the effects of different sizes of superconducting
energy gaps on the different bands, only a single isotropic gap is
associated with each pocket.  In order to reproduce the data in Fig.~\ref{fig:bcs}(a),
different combinations were considered.  The best choice is a linear combination
of the large gap ($2\Delta_2$) applied to the narrow Drude response in
Fig.~\ref{fig:bcs}(b) and the small ($2\Delta_1$) gap applied to the broad Drude
response in Fig~\ref{fig:bcs}(c), indicated by the shaded regions; this line has been
smoothed and is shown as the dash-dot line in Fig.~\ref{fig:bcs}(a), which manages
to reproduce the data quite well.  This is somewhat surprising for two reasons.
First, the curve has not been refined in any way, and second, this is a simple
superposition of two single-band BCS models and not a more sophisticated two-band
model of superconductivity that considers both intraband as well as interband
pairing \cite{maksimov11a,maksimov11b,karakozov14}.  On the other hand, since this
approach appears to work rather well, we speculate that the large difference in
the scattering rates in the two bands allows for this simpler interpretation.

%
%
Taking the contributions for the superconducting plasma frequencies from the two
bands,  $\omega_{p,S;1} \simeq 2300$~cm$^{-1}$ from the narrow band and
$\omega_{p,S;2} \simeq 2740$~cm$^{-1}$ from the broad band; the strength of
the condensate may be estimated by adding the two in quadrature, $\omega_{p,S}^2 =
\omega_{p,S;1}^2 + \omega_{p,S;2}^2$, yielding $\omega_{p,S} \simeq 3570$~cm$^{-1}$,
only somewhat larger than the experimentally-determined value of $\omega_{p,S} \simeq
3280\pm 200$~cm$^{-1}$.

%
%
The observation of two gap features is consistent with a number of recent
theoretical works that propose that isotropic \emph{s}-wave gaps form on
the electron and hole pockets but change sign between different Fermi
surfaces \cite{mazin08,chubukov09}, the so-called $s^{\pm}$ model.
However, there is considerable flexibility in this approach that allows
for situations in which the sign does not change between the
Fermi surfaces ($s^{++}$), $s^{\pm}$ with nodes on the electron pockets for
moderate electron doping, nodeless \emph{d}-wave superconductivity for
strong electron doping, as well as nodal \emph{d}-wave superconductivity
for strong hole doping \cite{chubukov12}.  The observation of multiple
gaps is also consistent with an ARPES study on this material which
observed isotropic gaps on all Fermi surfaces, with $\Delta_1 \simeq
2.5$~meV (hole pocket) and $\Delta_2 \simeq 4.2$~meV (electron pocket) \cite{miao12}.
These results are in reasonable agreement with the values determined
using our simple model, $\Delta_1 \simeq 2.8$~meV and $\Delta_2 \simeq
5.6$~meV, and the reduction of the conductivity at low frequency for
$T \ll T_c$ suggests the absence of nodes.  The ARPES study would tend
to suggest that the large gap associated with the electron pocket
corresponds to the weak, narrow Drude contribution, while the small
gap associated with the hole pocket corresponds to the strong, broad
Drude response.  This is also consistent with our earlier observation
of a relatively small scattering rate on the electron pocket in
Fe$_{1.03}$Te \cite{dai14}.
%

%
%
\subsection{Parameter scaling}
\begin{figure}[t]
%
%
%
\centerline{\includegraphics[width=0.90\columnwidth]{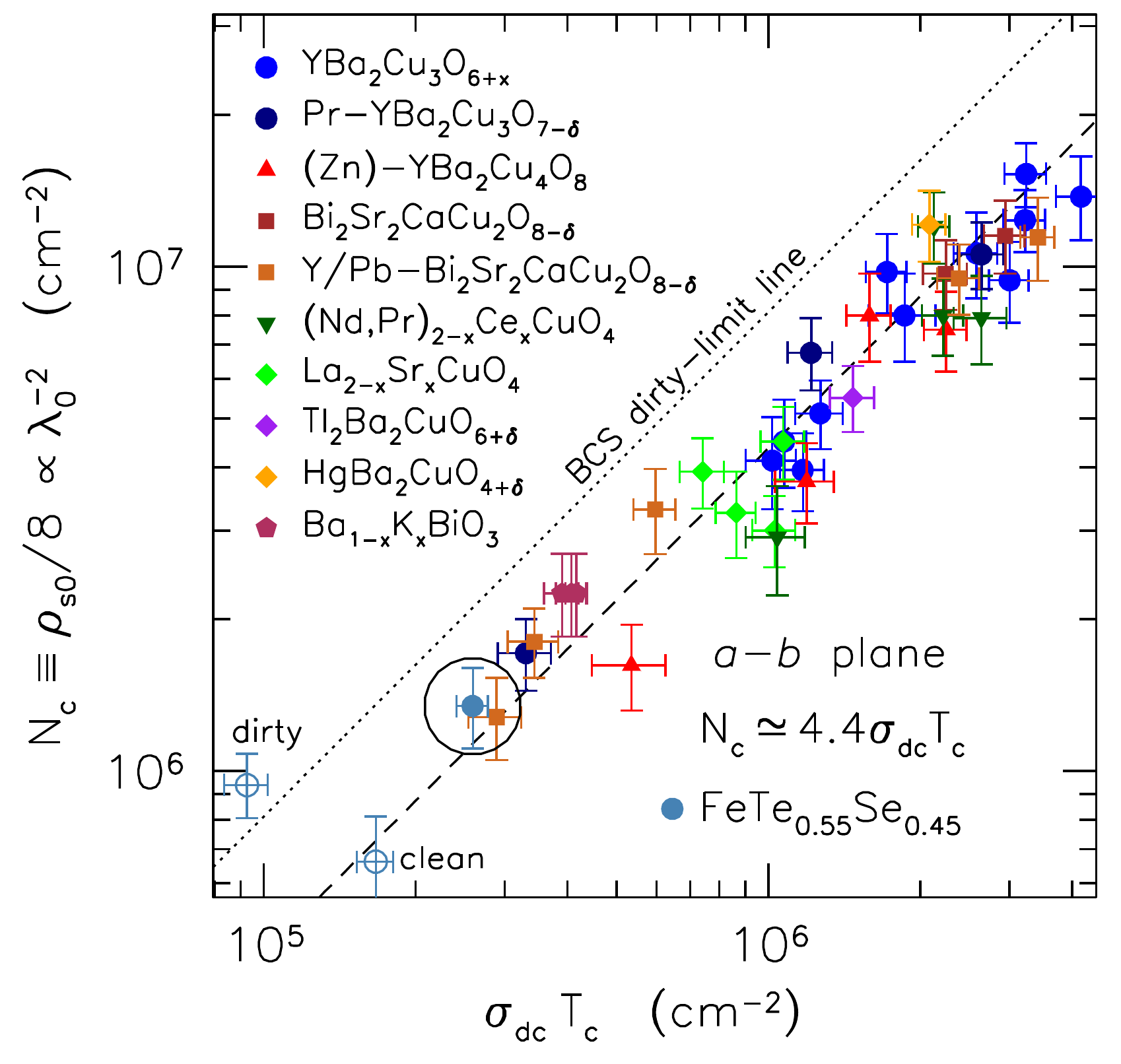}}%
\caption{(Color online) The log-log plot of the in-plane spectral
 weight of the superfluid density $N_c \equiv \rho_{s0}/8$ vs
 $\sigma_{dc}\,T_c$, for a variety of  electron and hole-doped
 cuprates compared with the result for FeTe$_{0.55}$Se$_{0.45}$.
 The dashed line corresponds to the general result for the cuprates
 $\rho_{s0}/8 \simeq 4.4 \sigma_{dc}T_c$, while the dotted line
 is the result expected for a BCS dirty-limit superconductor in
 the weak-coupling limit, $\rho_{s0}/8 \simeq 8.1\, \sigma_{dc}T_c$.
 The open circles represent the different contributions to the
 superfluid density in FeTe$_{0.55}$Se$_{0.45}$; the solid circle is
 the experimental value.}
\label{fig:scaling}
\end{figure}

In our previous study of this material, we noted that it fell on the general
scaling line originally observed for the high-temperature superconductors \cite{homes04,homes05},
recently demonstrated for some of the iron-based materials \cite{wu10b},
$\rho_{s0}/8\simeq 4.4\,\sigma_{dc}T_c$, where $\sigma_{dc}$ is measured just above
$T_c$.  A natural consequence of the BCS theory in the dirty limit is the emergence
of a similar scaling line \cite{homes05,tallon06} $\rho_{s0}/8 \simeq 8.1 \sigma_{dc} T_c$
(dotted line in Fig.~\ref{fig:scaling}).
%
%
%
%
The experimentally-determined values of $\sigma_{dc}\equiv\sigma_1(\omega\rightarrow 0)
= 5600\pm 400$~$\Omega^{-1}{\rm cm}^{-1}$ and $\omega_{p,S} \simeq 3300$~cm$^{-1}$
($\rho_{s0}\equiv \omega_{p,S}^2$) indicate that this material falls on the scaling
line in the vicinity of the underdoped cuprates, as shown in Fig.~\ref{fig:scaling}.
The decomposition of the superconducting response into two bands allows the different
contributions to the superfluid density to be examined (Fig.~\ref{fig:bcs}).
The dirty-limit contribution ($\sigma_{dc}\simeq 2000$~$\Omega^{-1}$cm$^{-1}$ and
$\omega_{p,S} \simeq 2740$~cm$^{-1}$) falls very close to the calculated BCS dirty-limit
scaling line, while the clean-limit contribution ($\sigma_{dc}\simeq 3600$~$\Omega^{-1}$cm$^{-1}$
and $\omega_{p,S} \simeq 2300$~cm$^{-1}$) falls to the right; this latter behavior
is expected and has been previously discussed \cite{homes05}.
Initially, it was thought that the materials that fell on the scaling line were
likely in the dirty limit \cite{homes05}.  However, it has been shown that many
superconducting materials fall on the scaling line, and many of them are not in
the dirty limit \cite{dordevic13}.  Moreover, it has been recently demonstrated
that the scaling relation is more robust than originally thought and should be
valid for most materials, including those that approach the clean limit \cite{kogan13},
suggesting that the scaling relation is an intrinsic property of the BCS theory of
superconductivity.  Therefore, even though the contributions to the superfluid density
in FeTe$_{0.55}$Se$_{0.45}$ come from the clean as well as the dirty limit,
the material should, and indeed does, fall on the universal scaling line.

%
%
\vspace*{4.0mm}
\section{Conclusions}
The detailed optical properties of the multiband superconductor FeTe$_{0.55}$Se$_{0.45}$
($T_c = 14$~K) have been examined for light polarized in the Fe-Te/Se (\emph{a-b})
planes for numerous temperatures above $T_c$, as well as several below.  In recognition
of the multiband nature of this material, the optical properties are described by the
two-Drude model.  In the normal state the two-Drude model
yields a relatively weak Drude response ($\omega_{p,D;1} \simeq 3000$~cm$^{-1}$) that
is quite narrow at low temperature ($1/\tau_{D,1} \simeq 30$~cm$^{-1}$ at 20~K) but
which grows quickly with increasing temperature, and a strong Drude response
($\omega_{p,D;1} \simeq 14\,500$~cm$^{-1}$) with a large scattering rate
($1/\tau_{D,2} \simeq 1420$~cm$^{-1}$) that is essentially temperature independent.
It is demonstrated that the generalized-Drude model may not be used reliably in multiband
materials, except in those cases where chemical substitution has effectively rendered the
material either completely electron- or hole-doped.
In the superconducting state for $T\ll T_c$ the optical conductivity is reproduced quite
well using the normal-state properties for $T\gtrsim T_c$ and Mattis-Bardeen formalism
with a small gap ($\Delta_1 \simeq 23$~cm$^{-1}$ or about 2.8~meV) applied to the strong
Drude component, and a large gap ($\Delta_2 \simeq 45$~cm$^{-1}$ or about 5.6~meV)
applied to the narrow Drude component.  Because the scattering rates on the two bands are quite
different, this places one band in the dirty limit ($1/\tau \gg \Delta$) and the
other close to the clean limit ($1/\tau \lesssim \Delta$), effectively placing this material
simultaneously in both the clean and dirty limit.  The estimate for the superfluid
density of $\rho_{s0} \simeq 3600$~cm$^{-1}$ using this model  is quite close
to the experimentally-determined value $\rho_{s0} \simeq 3300$~cm$^{-1}$, which places
this material on the universal scaling line for high-temperature superconductors in the
region of the underdoped cuprates, similar to other iron-based superconductors.
%
%
\vspace*{-2.0mm}
\begin{acknowledgements}
We would like to acknowledge illuminating discussions with L. Benfatto, K. Burch,
A. V. Chubukov, J. C. Davis, H. Ding, M. Dressel, Z. W. Lin , H. Miao and S. Uchida.
This work is supported by the Office of Science, U.S. Department of Energy
under Contract No. DE-SC0012704.
\end{acknowledgements}

%
%
%
%
%

\end{document}